\documentstyle [12pt]{article}
\begin{document}
\centerline {Rencontre de Moriond on "Dark Matter in Cosmology, Clocks and}
\centerline {Tests of Fundamental Laws", Villars (Switzerland),
January 21-28 1995}
\vskip 1cm
\centerline {\bf PROPERTIES OF}
\vskip 3mm
\centerline {\bf A POSSIBLE CLASS OF PARTICLES}
\vskip 3mm
\centerline {\bf ABLE TO TRAVEL}
\vskip 3mm
\centerline {\bf FASTER THAN LIGHT}
\vskip 2cm
\centerline {\bf L. GONZALEZ-MESTRES}
\vskip 4mm
\centerline {Laboratoire d'Annecy-le-Vieux de Physique des Particules,}
\centerline {B.P. 110 , 74941 Annecy-le-Vieux Cedex, France}
\vskip 1mm
\centerline {and}
\vskip 1mm
\centerline {Laboratoire de Physique Corpusculaire, Coll\`ege de France,}
\centerline {11 pl. Marcellin-Berthelot, 75231 Paris Cedex 05 , France}
\vskip 2cm
{\bf Abstract}
\vskip 2mm
The apparent Lorentz invariance of the laws of physics does not imply that
space-time is indeed minkowskian.
Matter made of solutions of Lorentz-invariant equations would feel
a relativistic space-time even if the actual space-time
had a quite different geometry (f.i. a galilean space-time).
A typical example is provided by sine-Gordon solitons
in a galilean world.
A "sub-world" restricted to such solitons would be "relativistic",
with the critical speed of solitons playing the
role of the speed of light.
Only the study of the deep structure of matter will unravel
the actual geometry of
space and time, which we expect to be scale-dependent and
determined by the properties of matter itself.
\par
If Lorentz invariance is a property of equations describing a
sector of matter at a given scale, an absolute frame
(the "vacuum rest frame") may exist without contradicting the
minkowskian structure of the space-time
felt by ordinary particles. But $c$ , the speed of light,
will not necessarily be the only critical speed in vacuum:
for instance, a superluminal sector of matter may exist
related to new degrees of freedom
not yet discovered experimentally. Such particles would not be tachyons:
they may feel a different minkowskian
space-time with a critical speed $c_1~>~c$
and behave kinematically like ordinary particles apart from the
difference in critical speed.
At $v$ (speed) $>~c$ , they are expected to release
"Cherenkov" radiation (ordinary particles) in vacuum.
We present a discussion of possible physical (theoretical and experimental)
and cosmological implications of such a scenario,
assuming that the superluminal sector couples weakly to ordinary matter.
\vskip 2cm
{\bf 1. RELATIVITY, SPACE-TIME AND MATTER}
\vskip 6mm
In textbook special relativity, mikowskian geometry
is an intrinsic property of space and time: any material body moves
inside a minkowskian space-time governed by Lorentz
transformations and relativistic kinematics. The action itself is
basically given by
the metrics. General relativity includes gravitation
and local invariance within this framework, but the "absoluteness" of the
previous concepts remains even
if matter modifies the local structure of space and time.
Gravitation is given a geometric description within the
Minkowskian approach: geometry remains the basic principle of
the theory and provides the ultimate dynamical concept.
Such an approach has widely influenced modern theoretical
physics and, especially, recent grand unified theories.
\vskip 4mm
On the other hand, a look to various dynamical systems studied
in the last decades would suggest a more flexible view of
the relation between matter and space-time.
Lorentz invariance can be viewed as a symmetry of the motion
equations, in which case no reference to absolute
properties of space and time is required and the properties of matter
play the main role. In a two-dimensional galilean space-time,
the equation:
\equation
\alpha ~\partial ^2\phi /\partial t^2~-~\partial ^2\phi /\partial x^2 = F(\phi
)
\endequation
with $\alpha$ = $1/c_o^2$ and $c_o$ = critical speed remains
unchanged under "Lorentz" transformations leaving invariant the square
interval:
\equation
ds^2 = dx^2 - c_o^2 dt^2
\endequation
so that matter made with solutions of equation (1) would feel a
relativistic space-time even if the real space-time is actually
galilean and if an absolute rest frame exists in the
underlying dynamics beyond the wave equation.
A well-known exemple is provided by the solitons of the
sine-Gordon equation taking in (1):
\equation
F(\phi ) = (\omega /c_o^2)~sin~\phi
\endequation
\par
A two-dimensional universe made of sine-Gordon solitons plunged
in a galilean world would behave like a two-dimensional minkowskian
world with the laws of special relativity.
Information on any absolute rest frame would be lost by the solitons.
\vskip 4mm
1-soliton solutions of the sine-Gordon equation are known to exhibit
"relativistic" particle properties. With $|v|$ $<$ $c_o$ ,
a soliton of speed $v$ is described by the expression:
\equation
\phi _v (x,t) =
4~arc~tan~[exp~(\pm~\omega c_o^{-1}~(x-vt)~(1- v^2/c_o^2)^{-1/2})]
\endequation
with the following properties:
\vskip 3mm
- size $\Delta x = c_o\omega ^{-1}~(1- v^2/c_o^2)^{1/2}$
\vskip 3mm
- proper time $d\tau = dt~(1- v^2/c_o^2)^{1/2}$
\vskip 3mm
- energy $E = E_o~(1- v^2/c_o^2)^{-1/2}$ , $E_o$ being the energy at rest
and $m = E_o/c_o^2$ the "mass" of the soliton
\vskip 3mm
- momentum $p = mv~(1- v^2/c_o^2)^{-1/2}$
\vskip 4mm
\noindent
so that everything looks perfectly "minkowskian" even
if the basic equation derives from a galilean world with an
absolute rest frame. In such case, the actual structure
of space and time can only be found by going beyond the wave equation
to deeper levels of resolution, similar to the way high energy accelerator
experiments explore the inner structure of
"elementary" particles.
The answer may then be scale-dependent and matter-dependent.
\vskip 4mm
Free particles move in vacuum, which is known
(i.e. from the Weinberg-Glashow-Salam theory) to be a material medium
where condensates and other structures can develop. We measure
particles with devices made of particles.
We are ourselves made of particles, and we are inside the vacuum.
All known particles have a critical speed in vacuum equal
to the speed of light, $c$ . But a crucial question  remains open:
is $c$ the only critical speed in vacuum,
are there particles with a critical speed different from that of light?
The question clearly makes sense, as in a
perfectly transparent crystal it is possible to
identify at least two critical speeds: the speed of light and
the speed of sound. The present paper is devoted to
explore a simple nontrivial scenario, with two critical speeds in vacuum.
\vskip 6mm
{\bf 2. PARTICLES IN VACUUM}
\vskip 5mm
Free particles in vacuum usually satisfy a dalembertian equation,
such as the Klein-Gordon equation for scalar particles:
\equation
(c^{-2}~\partial ^2/\partial t^2~-~\Delta)~\phi~~+~~m^2c^2~(h/2\pi
)^{-2} \phi ~~=~~0
\endequation
where the coefficient of the second time derivative sets $c$ ,
the critical speed (speed of light), and, given $c$ and the Planck
constant $h$ , the coefficient of the linear term in $\phi $ sets $m$ ,
the mass of the particle.
To build plane wave solutions, we consider the following physical
quantities given by differential operators:
\vskip 3mm
\centerline {$E = i~(h/2\pi )~\partial /\partial t$~~~~~~,~~~~~~
$\vec{\bf p} = -i~(h/2\pi )~\vec{\bf \nabla } $}
\vskip 3mm
\noindent
and with the definitions:
\vskip 3mm
\centerline {$x^o = ct$~~~~~~,~~~~~~
$p^o = E/c$ ~~~~~~, ~~~~~~
$E = (c^2\vec{\bf p}^2~+~m^2c^4)^{1/2}$}
\vskip 3mm
\noindent
the plane wave is given by:
\equation
\phi (x,t) = exp~[-(2\pi i/h)~(p^ox^o-\vec{\bf p} . \vec{\bf x} )]
\endequation
\noindent
from which we can build position and speed operators [1]:
\equation
\vec{\bf x}_{op} =
(ih/2\pi )~(\vec{\bf \nabla }_p~-~(\vec{\bf p}^2+m^2c^2)^{-1}\vec{\bf p})
\endequation
\noindent
in momentum space, and:
\equation
\vec{\bf v} = d\vec{\bf x}_{op} /dt = (2\pi i/h)~[H ,
\vec{\bf x}_{op}] = (c/p_o)~\vec{\bf p}
\endequation
\noindent
where H is the hamiltonian and the brackets mean commutation. We then get:
\vskip 3mm
\centerline {$p_o = mc~(1~-~v^2/c^2)^{-1/2}~~~~~~,~~~~~~
\vec{\bf p} = m\vec{\bf v}~(1~-~v^2/c^2)^{-1/2}$}
\vskip 3mm
\noindent
and, at small $v/c$ :
\vskip 3mm
\centerline {$E_{free} \simeq 1/2~mv^2 ~~~~~~,~~~~~~
\vec{\bf p} \simeq m\vec{\bf v} $}
\vskip 3mm
\noindent
in which limit, taking $H = 1/2~mv^2~+~V(\vec{\bf x}_{op})$ , we obtain:
\vskip 3mm
\centerline {$\vec{\bf F} = - \vec{\bf \nabla }V = m~d\vec{\bf v} /dt$}
\vskip 3mm
\noindent
which shows that $m$ is indeed the inertial mass.
\vskip 4mm
A superluminal sector of matter can be consistently generated,
with the conservative choice of leaving the Planck constant
unchanged, replacing in the above construction the speed of light $c$
by a new critical speed $c_1$ $>$ $c$ . All
previous concepts and
formulas remain correct, leading to particles with
positive mass and energy which are not tachyons
and have nothing to do with previous proposals in this field [2].
The new particles
will have a larger rest energy, $E_{rest} = mc_1^2$ ,
for a given inertial mass.
To produce superluminal mass at accelerators may therefore require
very large amounts of energy.
In the "non-relativistic" limit $v/c_1$ $\ll $ 1 , kinetic
energy and momentum will remain given by the same expressions as before.
Energy and momentum conservation will in principle not be spoiled
by the existence of several critical speeds in vacuum.
\vskip 6mm
{\bf 3. A SCENARIO WITH TWO CRITICAL SPEEDS IN VACUUM}
\vskip 5mm
Assume a simple and schematic scenario, with two sectors of matter:
\vskip 3mm
- the "ordinary sector", made of "ordinary particles"
with a critical speed equal to the speed of light $c$ ;
\vskip 3mm
- a superluminal sector, where particles have a
critical speed $c_1$ $\gg $ $c$ .
\vskip 4mm
Several basic questions arise:
can the two sectors interact, and how? what would be the conceptual
and experimental consequences? can we observe
the superluminal sector and detect its particles? what would be the best
experimental approach? It is obviously impossible
to give general answers independent of the details of the scenario
(couplings, symmetries, parameters...),
but some properties and potentialities can be pointed out:
\vskip 4mm
- Even if each sector has its own "Lorentz invariance"
involving as the basic parameter the critical speed in vacuum of
its own particles, interactions between the two sectors
will break both Lorentz invariances. Even if the interaction is
mediated by scalar fiels preserving apparent Lorentz invariance
in the lagrangian density (e.g. with a $|\phi _o(x)|^2|\phi _1(x)|^2$
term where the complex scalar field $\phi _o$ belongs to
the ordinary sector and the complex scalar field $\phi _1$
to the superluminal one), the Fourier expansion of the scalar
fields shows the unavoidable breaking of both Lorentz invariances.
\vskip 3mm
- Even before considering interaction between the two sectors,
Lorentz invariance
for both sectors simultaneously will at best be explicit only in a single
inertial frame (the $vacuum$ $rest$ $frame$).
Apart from the trivial case of space rotations,
no linear space-time transformation can simultaneously preserve the
invariance of lagrangian densities for both sectors.
However, it will be impossible to identify the vacuum rest frame
if one of the sectors produces no measurable effect
(e.g. if superluminal particles and their influence on the ordinary
sector cannot be observed).
In our approach, the Michelson-Morley result is not necessarily
incompatible with the existence of some "ether" (the vacuum
as a material medium) clearly suggested by recent developments
in particle physics.
Finding some track of a superluminal sector (e.g.
through violations of Lorentz invariance in the ordinary sector) may
be the only way to experimentally discover the vacuum rest frame.
\vskip 3mm
- If superluminal particles couple to ordinary matter, they will
not in general be found traveling at a speed higher than $c$ (except
near the vertex of accelerator experiments). At
superluminal speed, such particles are expected to release "Cherenkov"
radiation (i.e. ordinary particles, whose emission in vacuum
is kinematically allowed in such case) until they will be decelerated
to a speed $v \leq c$ . In accelerator experiments, this
"Cherenkov" radiation may provide a clean signature to identify produced
superluminal particles.
Theoretical studies of tachyons rejected [3] the possibility of
"Cherenkov" radiation in vacuum because tachyons are not really
different from ordinary particles (they sit in a different kinematical
branch, but are the same kind of matter). However, in our case
we are dealing with a different kind of matter but superluminal
particles will always be in the region
of $E$ and $\vec{\bf p}$ real, with $E$ $=$
$(c_1\vec{\bf p}^2~+~m^2c_1^4)^{1/2} > 0$
and can emit "Cherenkov" radiation.
\vskip 3mm
- Gravitation is usually a gauge interaction, related to invariance
under local linear transformations of space-time and mediated by
a massles ordinary particle (the graviton) which
is expected to travel at $v$ = $c$ .
Since the graviton belongs to the ordinary sector, it is
not expected to play a universal role in the presence of a superluminal
sector. As each sector has its own Lorentz invariance, the
superluminal sector may generate its own "gravity" with a new
"graviton" traveling at critical speed $c_1$ and a new "Newton constant".
"Gravitational" interactions between the two sectors
(including "graviton" mixing) can be generated through the above
considered pair of complex scalar fields,
but this will lead to anomalies in "gravitational" forces for both sectors.
"Gravitational" coupling between ordinary and superluminal
particles is in this context expected to be weak. Concepts
so far considered as very fundamental (i.e. the universality of the
exact equivalence between inertial and gravitational mass)
will now fail and leave us with only approximate sectorial
properties, even if the real situation may be very difficult to
unravel experimentally.
\vskip 3mm
- Stability under radiative corrections (e.g. of the existence of
well-defined "ordinary" and "superluminal" sectors)
is not always ensured. As the critical speed is related to particle
properties in the region of very high energy and momentum,
the ultraviolet behaviour of the renormalized theory
(e.g. renormalized propagators) will be crucial.
However, work on supersymmetry, supergravity and other theories
suggests that technical solutions can be found to preserve
the identity of each sector
as well as the stability of the scheme.
\vskip 3mm
- Superluminal particles may have played a cosmological role
leading to substantial changes in the
"Big Bang" theory and to a reformulation of the problem of
the cosmological constant. Their annihilation into ordinary
particles may have generated expansion phenomena similar to
inflation, allowing to better describe the fomation of
large scale structure. Relic superluminal particles may exist,
and even dominate the Universe.
\vskip 6mm
{\bf 4. SOME PRACTICAL CONSIDERATIONS}
\vskip 5mm
Searching for effects indicating the possible existence of a
superluminal sector appears to be a difficult task. We present
here some preliminary remarks:
\vskip 4mm
- At accelerators, hadrons may be the best probe to produce
and observe superluminal particles as quarks
are coupled to all known interactions. Machines such as LHC have
thus interesting potentialities in the field, whereas $e^+e^-$
collisions should be preferred only if superluminal particles
couple to the electroweak sector. In an accelerator experiment,
a pair of superluminal particles would be produced at $E$
(available energy) $=~2mc_1^2$ and Cherenkov effect in vacuum
will start only slightly
above, at $E~=~2mc_1^2~+~mc^2$ $=$ $2mc_1^2 (1~+~1/2~c_1^2/c_2^2)$
$\simeq $ $2mc_1^2$ .
The Cherenkov cones will quickly become broad, leading to
"almost $4\pi $" events in the rest frame of the superluminal pair.
\vskip 3mm
- Effects of the superluminal sector on the ordinary one may be
basically high energy and short distance phenomena, far from
conventional tests of Lorentz invariance. Thus, nuclear and particle
physics experiments may open new windows. Apart from accelerator
 experiments, the search for abnormal effects in low energy nuclear
physics or in neutrino physics (with neutrinos moving close
to speed of light with respect to the vacuum rest frame)
deserves serious consideration.
\vskip 3mm
- The present density of superluminal particles, as well as their
gravitational properties, are fundamental and
unknown parameters in our scenario. Such particles may be part
of the dark matter and, if there is a large amount of
superluminal matter in the Universe, direct detection may be
possible in underground laboratories as well as through
pair annihilation in "astro-particle" experiments.
\vskip 3mm
- Although it seems normal to assume that the superluminal sector
is protected by a quantum number and that, at least, the
"lightest superluminal particle" will be stable, this is not
unavoidable and we may be inside a sea of very long-lived
superluminal particles which decay into ordinary particles.
Such decays may then be observable and even play a cosmological role.
\vskip 3mm
- Finally, it should be noticed that we have kept the value of the
Planck constant unchanged when building the superluminal sector.
This is not really an arbitrary choice, as conservation and
quantization of angular momentum make natural our hypothesis
if the superluminal and ordinary sector interact.
It seems justified to start the search for superluminal particles
assuming that their basic quantum properties
are not fundamentally different from those of ordinary particles.
\vskip 6mm
{\bf References}
\vskip 4mm
\noindent
[1] See, for instance, S.S. Schweber, "An Introduction to
Relativistic Quantum Field Theory". Row, Peterson and Co. 1961 .
\par
\noindent
[2] See, for instance, "Tachyons, Monopoles and Related Topics",
Ed. E. Recami. North-Holland 1978 .
\par
\noindent
[3] See, for instance, E. Recami in [2].
\end{document}